\documentclass[journal]{IEEEtran}
\usepackage{amsmath,amsfonts}
\usepackage{algorithmic}
\usepackage{algorithm}
\usepackage{array}
\usepackage[caption=false,font=normalsize,labelfont=sf,textfont=sf]{subfig}
\usepackage{textcomp}
\usepackage{stfloats}
\usepackage{url}
\usepackage{verbatim}
\usepackage{graphicx}
\usepackage{float}
\usepackage{color}
\usepackage{hyperref} 
\usepackage{multirow}
\usepackage{bbding}
\usepackage{pifont}
\usepackage{wasysym}
\usepackage{amssymb}
\usepackage{arydshln}
\usepackage{cite}
\usepackage{pdfpages}
\begin{document}

\title{
Accurate Lung Nodule Segmentation with \\ Detailed Representation Transfer and Soft Mask Supervision
}

\author{Changwei Wang,
        Rongtao Xu,
        Shibiao Xu~\IEEEmembership{Member,~IEEE,}
        Weiliang Meng~\IEEEmembership{Member,~IEEE,}
        \\
        Jun Xiao~\IEEEmembership{Member,~IEEE,}
       and Xiaopeng Zhang,~\IEEEmembership{Member,~IEEE}
        %and~Jane~Doe,~\IEEEmembership{Life~Fellow,~IEEE}% <-this % stops a space
\thanks{
C. Wang and R. Xu contributed equally. 
S. Xu and W. Meng are the corresponding authors (shibiaoxu@bupt.edu.cn; weiliang.meng@ia.ac.cn).}
\thanks{
C. Wang, R. Xu, W. Meng, J. Zhang and X. Zhang are with School of Artificial Intelligence, University of Chinese Academy of Sciences and National Laboratory of Pattern Recognition, Institute of Automation, Chinese Academy of Sciences. 
S. xu is with school of Artificial Intelligence, Beijing University of Posts and Telecommunications.
Jun Xiao is with School of Artificial Intelligence, University of Chinese Academy of Sciences.

}
}

\markboth{Journal of \LaTeX\ Class Files,~Vol.~XX, No.~XX, April~2022}%
{Shell \MakeLowercase{\textit{et al.}}: A Sample Article Using IEEEtran.cls for IEEE Journals}

\IEEEpubid{0000--0000/00\$00.00~\copyright~2021 IEEE}

\maketitle

\begin{abstract}
Accurate lung lesion segmentation from Computed Tomography (CT) images is crucial to the analysis and diagnosis of lung diseases such as COVID-19 and lung cancer. 
However, the smallness and variety of lung nodules and the lack of high-quality labeling make the accurate lung nodule segmentation difficult. 
To address these issues, we first introduce a novel segmentation mask named ``\textit{Soft Mask}" which has richer and more accurate edge details description and better visualization, and develop a universal automatic \textit{Soft Mask} annotation pipeline to deal with different datasets correspondingly.
Then, a novel \textbf{Net}work with detailed representation transfer and \textit{Soft Mask} supervision (\textbf{DSNet}) is proposed to process the input low-resolution images of lung nodules into high-quality segmentation results. 
Our DSNet contains a special Detail Representation Transfer Module (DRTM) for reconstructing the detailed representation to alleviate  the small size of lung nodules images, and an adversarial training framework with  \textit{Soft Mask} for further improving the accuracy of segmentation.
Extensive experiments validate that our DSNet outperforms other state-of-the-art methods for accurate lung nodule segmentation, and has strong generalization ability in other accurate medical segmentation tasks with competitive results.
Besides, we provide a new challenging lung nodules segmentation dataset for further studies. %[Dataset release]
\end{abstract}
\begin{IEEEkeywords}
Medical images segmentation, lung nodules segmentation, Soft Mask, detailed representation transfer.
\end{IEEEkeywords}

\section{Introduction}
\begin{figure}[t]
\begin{center}
  \includegraphics[width=1\linewidth]{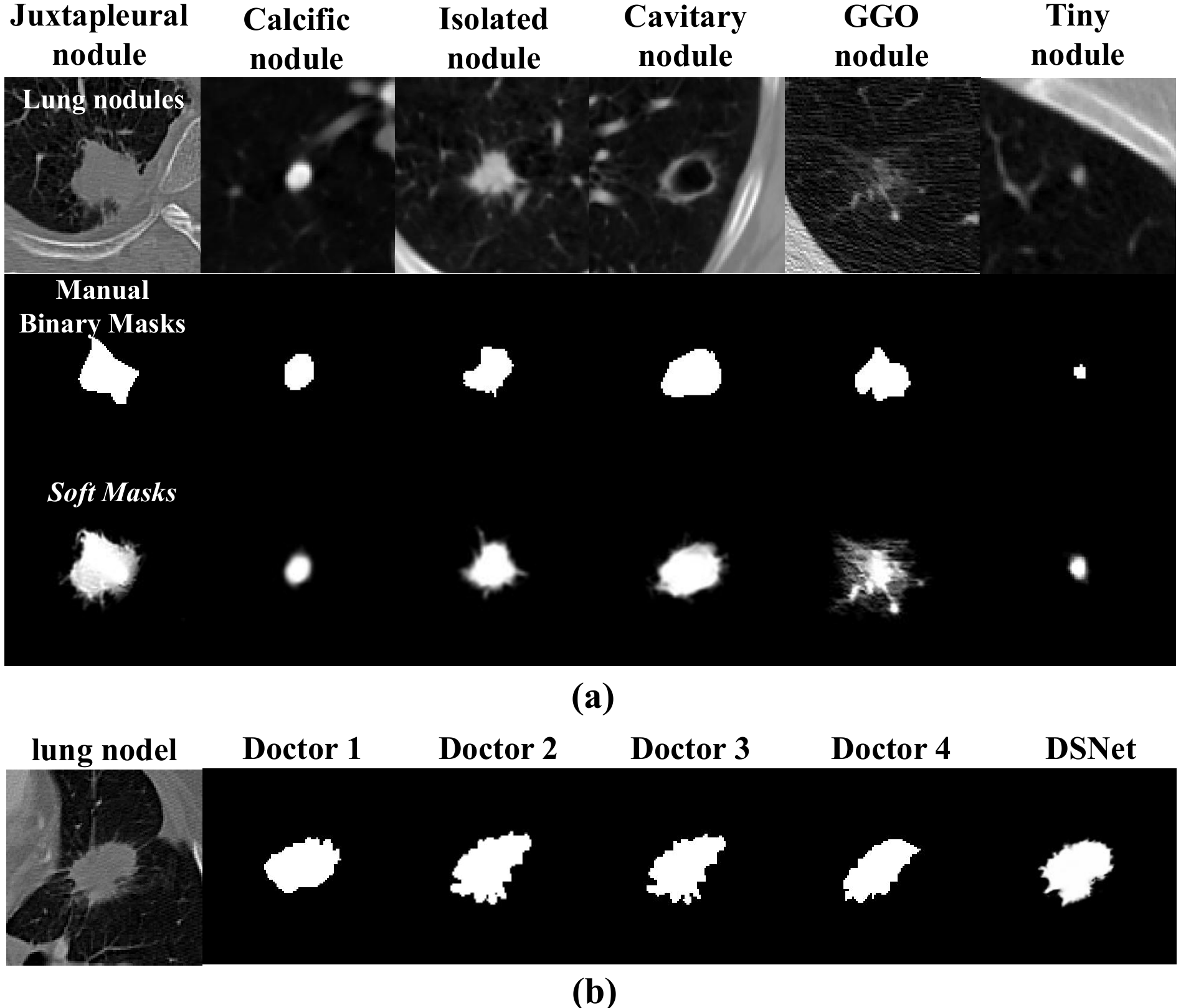}
\end{center}
   \caption{(a):Various types of lung nodules and manual labelings in LIDC.
   (b): Comparison of the labeling results of four doctors in LIDC with  prediction of our proposed DSNet. Our proposed DSNet even achieves better segmentation and visualization results than human radiologists.
  }
\label{fig:introduction}
\end{figure}

\IEEEPARstart{S}{ince} December 2019, the world has suffered a severe health crisis: COVID-19 pandemic~\cite{wang2020novel}. COVID-19 usually appears as a ground glass opacity (GGO) lung nodule on computed tomography (CT) images~\cite{wang2021focus}.
These GGO and other types of lung nodules potentially contain the risk of lung cancer, which is the deadliest type of cancer worldwide for human with a relatively low five-year survival rate of 18\%~\cite{siegel2019cancer}. Accurate segmentation of the lung nodule is of great significance for automated disease screening, diagnosis, analysis, and treatment evaluation~\cite{macmahon2005guidelines,liao2019evaluate}.
However, accurate lung nodule segmentation is a challenging problem due to the following various reasons. 

\IEEEpubidadjcol
“One cannot make bricks without straw” is a commonly used expression to imply that it will be challenging to complete the work without the support of the primary  conditions. Unfortunately, this tends to be the case for the current accurate segmentation of lung nodules without high-quality mask annotations.
High-quality mask annotation datasets for accurate segmentation of lung nodules are absent, because 
manually labeling lung nodules mask is a time-consuming work obviously~\cite{pedrosa2019lndb}, while lung nodules annotation by radiologists is a highly subjective task, often influenced by individual bias and clinical experiences. As shown in Fig.~\ref{fig:introduction} (b), there are apparent differences in the labels of four doctors, and these inconsistencies may cause ambiguities in the segmentation network training.  Cai et al.~\cite{cai2018accurate} proposed an automatic mask generation method using the long and short axis of RECIST~\cite{eisenhauer2009new} as weakly prior information to avoid manually label the mask, while this is an iterative process, and can only obtain a rough mask by relying on coarse RECIST marks and GrabCut~\cite{grabcut}, making it difficult to cope with the task of accurate segmentation of lung nodules with complex edges. 

In addition, some inherent characteristics of lung nodules also cause difficulty in their segmentation. On the one hand, lung nodules are usually small (5 $ \sim $ 10 mm)~\cite{macmahon2005guidelines}, which leads to the low resolution of the input image. Obviously, low-resolution input loses detailed information of the lesion and makes it difficult to be segmented accurately. On the other hand, as shown in Fig.~\ref{fig:introduction} (a), lung nodules have significant heterogeneity. 
Some lung nodules such as GGO and burr nodules have irregular shapes and complex edges which also pose challenges for accurate segmentation~\cite{lassen2015robust}. However, as shown in Fig.~\ref{fig:introduction}, the existing binary mask cannot describe the edge details of these nodules well, so it is difficult to achieve accurate segmentation and visualization in this case.

%Second, Lung nodules have significant heterogeneity, in which such as GGO and juxtapleural nodules have irregular shapes and tiny edges shown in Fig.~\ref{fig:introduction}, bringing challenges for accurate segmentation, while existing binary masks cannot describe the edge details of these nodules well. Some lung nodules such as GGO and burr nodules have irregular shapes and complex edges which also pose challenges for accurate segmentation~\cite{lassen2015robust}. However, as shown in Fig.~\ref{fig:introduction} (a) and (b), the existing binary mask cannot describe the edge details of these nodules well, so it is difficult to achieve accurate segmentation and visualization in this case. In addition, some inherent characteristics of lung nodules also cause difficulty in their segmentation. Little lung nodules are usually small (5 $ \sim $ 10 mm)~\cite{macmahon2005guidelines}, which leads to the low resolution of the input image. Obviously, low-resolution input image loses detailed information about the lesion and makes it difficult to be segmented accurately. %Besides, as shown in Fig.~\ref{fig:introduction} (a), lung nodules have significant heterogeneity. 

To address the above issues, we propose an innovative paradigm including a automatic accurate annotation pipeline and a segmentation network named \textbf{DSNet}  with detailed representation transfer and \textit{Soft Mask} supervision for accurate lung nodules segmentation. 
Superior to most deep learning-based lung nodules segmentation methods~\cite{cfnet,2017Automatic,2019A,dual}, we not only improve the segmentation network architecture,  but also improve the qualities of both the less-explored input (low-resolution images) and ground truth masks(rough binary masks).
Specifically, our \textit{Soft Mask} can preserve rich edge details and smooth transition between lung nodules and surrounding pathological environments, making it have a  better visualization effect and richer information. 
And we develop an automatic accurate annotation pipeline to derive accurate \textit{Soft Masks} from different datasets. 
In addition, inspired by the recently popular self-supervised pre-training model~\cite{moco}, we carefully design a Detailed Representation Transfer Module (DRTM) in DSNet to transfer detailed representation knowledge from an off-the-shelf SRGAN~\cite{srgan} model. It can alleviate the problem of the low resolution of the input images.
In general, our technical contributions have the following four aspects:
\begin{itemize}
\item We introduce a new fine mask form ``\textit{Soft Mask}" and the automatic accurate labeling pipeline, which can boost segmentation accuracy and obtain excellent visualization in contrast to the traditional binary mask.
 
\item We design a novel DSNet with a Detailed Representation Transfer Module and a \textit{Soft Mask}
based adversarial training framework to convert low-quality images input into high-quality segmentation results. 
Unlike most related methods, we improve the quality of the input image and supervision mask that were rarely explored before.
 
\item We propose a special accurate lung nodules segmentation paradigm based on the above innovation, and its performance ourperforms other state-of-the-art methods on lung nodules segmentation.    
In addition, it is also suitable for other accurate medical segmentation tasks after our validation.

\item We provide a new lung nodules segmentation dataset (\textit{LNSM}\footnote{It is available at \href{https://drive.google.com/file/d/15NNkvDTb_0Ku0IoPsNMHezJRTH1Oi1wm/view?usp=sharing}{[dataset release]}.}) with both \textit{Soft Mask} labelings and binary masks, containing $1500$ accurate segmentation nodules. 
\end{itemize}

This work extends our preliminary conference version~\cite{softgan}~\footnote{It was accepted as oral by ICME 2022 and see the supplementary material for full paper.} in the following aspects. 
\textbf{Firstly}, we provide a general effective medical super-resolution model based on our self-supervised training and apply its detailed representation to promote medical segmentation task (Sec.~\ref{sec: drt} (1)).
\textbf{Secondly}, we design a novel Detailed Representation Transfer Module to flexibly transfer the detailed information of the super-resolution model into the segmentation pipeline (Sec.~\ref{sec: drt} (2)).
\textbf{Thirdly}, we increase experimental validation of our method's robustness and generalization.
Specifically, we validate the robustness by comparing models trained on cross-domain datasets with human doctors ( Sec.~\ref{sec: expr}), then apply to medical segmentation datasets other than lung nodule segmentation to verify the generalization of our method (Sec.~\ref{sec: expg}).
\textbf{Finally}, we also provide a more inclusive and insightful
discussion on our method (see Sec.~\ref{sec: discussion}).

% Related Work
\section{Related Work}
\subsection{Lung Nodules Segmentation}
Many classic hand-craft features based methods have been proposed to deal with lung nodules segmentation, such as morphological operations based methods~\cite{kostis2003three}, region-growing methods~\cite{kubota2011segmentation}, energy optimization based method~\cite{chan2001active,boykov2004experimental}, Conditional Random Field (CRF) based method~\cite{wu2010stratified}. However, these methods cannot cope with lung nodules segmentation well~\cite{dual}, especially for irregular-shaped nodules.

In recent years, deep learning based methods have gained new attention for medical image analysis and processing. Some convolutional neural networks (CNN) based lung nodules segmentation methods~\cite{cfnet,2017Automatic,2019A,dual} have achieved highly competitive segmentation accuracy. In addition, some other medical segmentation methods~\cite{unet,unet++,infnet,tomar2022fanet,paluru2021anam,wang2021focus} have brought breakthroughs in a variety of medical segmentation tasks. More recently, some transformer-based models~\cite{medt,transunet} are proposed for medical image segmentation.
These methods employ more complex network architecture to improve the segmentation accuracy, such as the use of attention mechanisms, spatial context, and dense connections. In contrast, our DSNet not only improves the network structure but also creatively ameliorates the quality and detail of both input images and supervised ground truth to achieve accurate lung nodules segmentation and impressive visualization.

\subsection{Soft Labels}
Soft labels have recently been applied to brain lesions segmentation~\cite{softlabel,softseg} as they are considered to have a better generalization, faster learning speed, and mitigation of network over-confidence.  Specifically, Kats et al.~\cite{softlabel} employed morphological dilation to expand the binary mask and assigned a fixed probability to all pixels within the expanded region to generate soft labels. Gros et al.~\cite{softseg}  obtained soft labels by bilinear interpolation, while they still lost many edge details (e.g. small burrs around lung nodules). Both the soft labels in~\cite{softlabel} and ~\cite{softseg} softened the binary masks, but they are still too rough to obtain accurate lung nodule segmentation with good visual effects. In contrast, the pixels of our \textit{Soft Mask} are not discrete but continuous, meaning that our \textit{Soft Mask} has richer and more accurate edge detail expression which can reduce the impact of imprecise boundary annotation. Our \textit{Soft Mask} can be obtained from various 
datasets. In particular, the RECIST marks in the current hospital picture and archiving systems (PACS) can be employed for quick generation of large-scale and high-quality \textit{Soft Masks}.

\subsection{Detailed Representation in Medical images}
High-resolution medical images are usually preferred in clinical practice due to more clear image structure and texture details, as well as the benefits to subsequent analysis and processing~\cite{sr}. These detailed representations in high-resolution images are very important for accurate segmentation, especially in edge regions.  Unfortunately, the extremely small lung nodules are difficult to overcome the challenges of hardware, physical and physiological factors to obtain high-resolution images under the existing imaging system. In recent years, some deep learning-based natural images or medical images super-resolution methods~\cite{srgan,sr1,sr2,you2022fine} have excellent performance in image resolution improvement. However, the lack of high-resolution training data limits the application of these methods. In addition, the potential of combining these methods with medical segmentation has not been fully exploited. In this work, we conduct self-supervised training on a super-resolution model for medical images and smartly transfer the detailed representation knowledge in the trained super-resolution model to the lung nodule segmentation pipeline.

% Method
\section{METHODOLOGY}
In this section, we propose a complete paradigm for accurate lung nodule segmentation. 
First, we introduce the \textit{Soft Mask} and develop a pipeline to label it automatically. Then, we design a novel DSNet with detailed representation transfer and \textit{Soft Mask} based adversarial training framework. 
%Finally, we provide the implementation details.

\subsection{\textit{Soft Mask} for Lung Nodule Segmentation}
%\vspace{0.2cm}
\subsubsection{Definition of \textit{Soft Mask}}
\label{sec: define}
Lung nodules are usually labeled and segmented in the form of a binary mask in most methods. 
However, the binary mask has many disadvantages in accurately segmenting lung nodules.
First, it cannot clearly describe the edge and morphological details of the lung nodules, which may cause anatomical information in these marginal regions to be ignored. Second, the labeling is extremely unbalanced since most of the regions are labeled as non-lesion in the binary mask~\cite{softlabel}, and this imbalance impairs the training of the segmentation network. 

\begin{figure}[t]
\begin{center}
  \includegraphics[width=1 \linewidth]{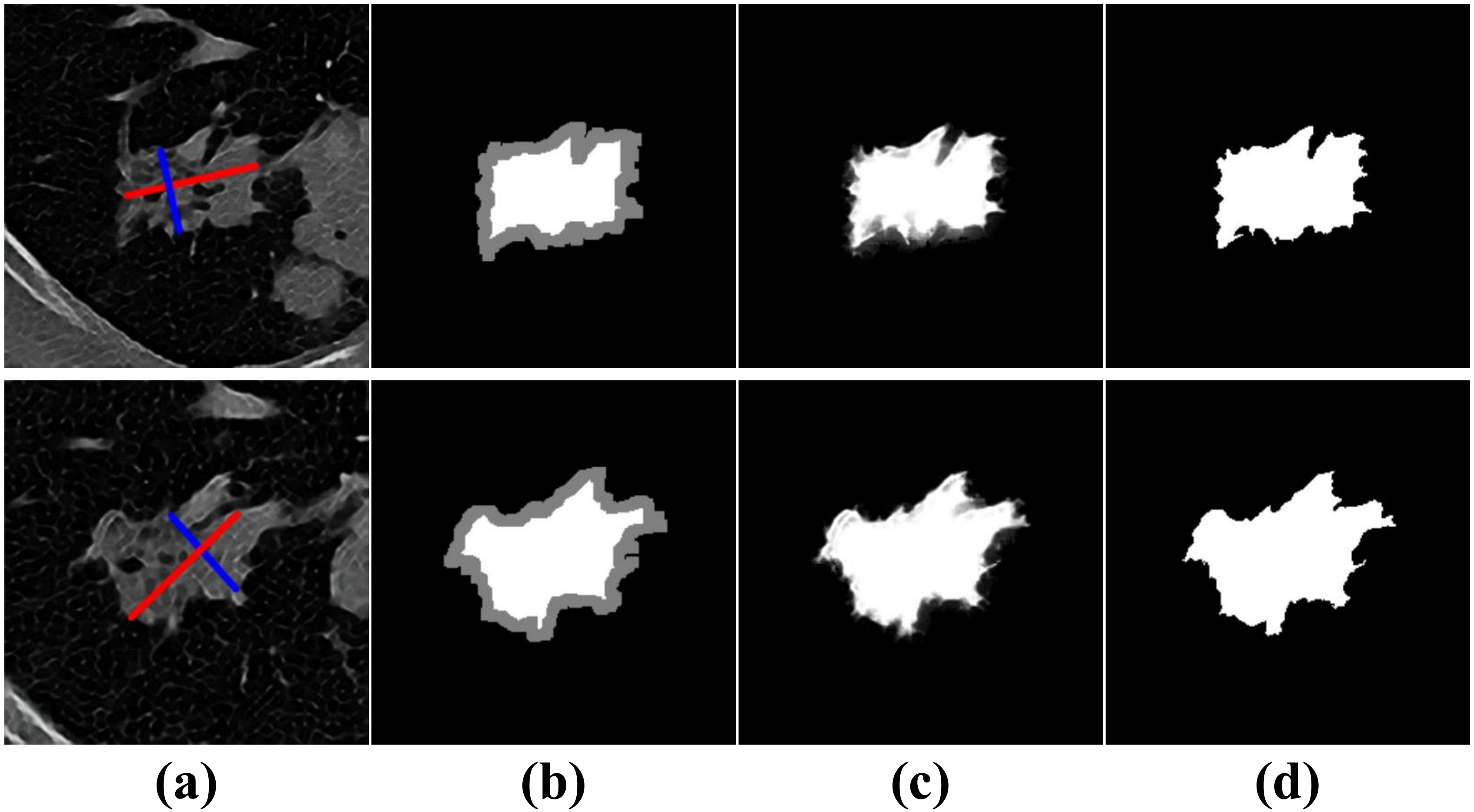}
\end{center}
   \caption{(a) Lung nodules with RECIST marks. The {\color{red}{\bf red line}} is the long axis and the {\color{blue}{\bf blue line}} is the short axis. (b) Trimaps.  \textbf{White pixels} mean the lesion region. \textbf{Black pixels} mean the background region.   \textbf{Gray pixels} mean the uncertain region. (c) \textit{Soft Masks}. (d) Binarized \textit{Soft Masks}. The \textit{Soft Mask} preserves edge details of lung nodules effectively.}
\label{fig:mask}
\end{figure}

To overcome the weakness of the binary mask, we introduce a new form of accurate labeling called \textit{Soft Mask} for lung nodule segmentation inspired by the image matting task~\cite{zheng2022image}.
We define the $i$-th pixel of \textit{Soft Mask} $M_i$  as a linear combination of the lesion label $L_i$ (the value is 1) and non-lesion background label $B_i$ (the value is 0):
\begin{align}
M_{i}=\alpha _{i}L_{i}+(1-\alpha _{i})B_{i},
\label{eq: eq1}
\end{align}
where $\alpha$ means the extent of belonging to the lung nodules region.
Here the close-form-matting algorithm~\cite{cfm} is employed to solve the $\alpha$ matrix, which converts the problem into a closed-form solution through linear assumptions and the labeling pixels. 
The labeling pixels are the key prior information for solving the above problems and are represented by trimap, which consists of lesion pixels, background pixels, and uncertain area pixels shown in Fig.~\ref{fig:mask} (b). 
Therefore, it is important to obtain a high-quality trimap.

%\vspace{-0.2cm}

\subsubsection{Labeling of \textit{Soft Mask}}
\label{sec label}
We develop an automatic \textit{Soft Mask} labeling pipeline to deal with priors from different datasets. 

First, we initialize these different priors into trimaps composed of lesion region, background region, and the uncertain region as shown in Fig.~\ref{fig:mask}~(b).
Specifically, we design the following three strategies for the trimap initialization:

\noindent\textbf{(1) Trimap Generation with the normal binary mask:} We use morphological operations to process the binary mask. Specifically, The lesion region is obtained by the erosion operation, while the non-lesion background region is obtained by the dilation operation, and the uncertain region is in the middle.

\noindent\textbf{(2) Trimap Generation with the binary mask of different doctors:} In order to improve the quality and reliability of labeling, some datasets are repeatedly labeled by different doctors. In LIDC dataset~\cite{lidc}, the mask for each lung nodule is marked by four different doctors, as shown in Fig.~\ref{fig:introduction}~(b). We set the intersection of these different masks labeled by doctors as the lesion region, while we set the complement of the union of these masks to the background region, and other pixels belong to the uncertain region.
Formally it can be expressed as:
\begin{align}
L= M_{d1}\cap M_{d2}\cap M_{d3} \cap  M_{d4},
\end{align}
\begin{align}
B=\overline{M_{d1}\cup M_{d2}\cup M_{d3} \cup  M_{d4}},
\end{align}
\begin{align}
U=\overline{L\cup B},
\end{align}
where $L$ denotes lesion region, $B$ denotes background region and $U$ denotes uncertain region, with $M_{di}$ representing the manual annotation of the $i$-th doctor.
\begin{figure*}[ht]
\begin{center}
  \includegraphics[width=1 \linewidth]{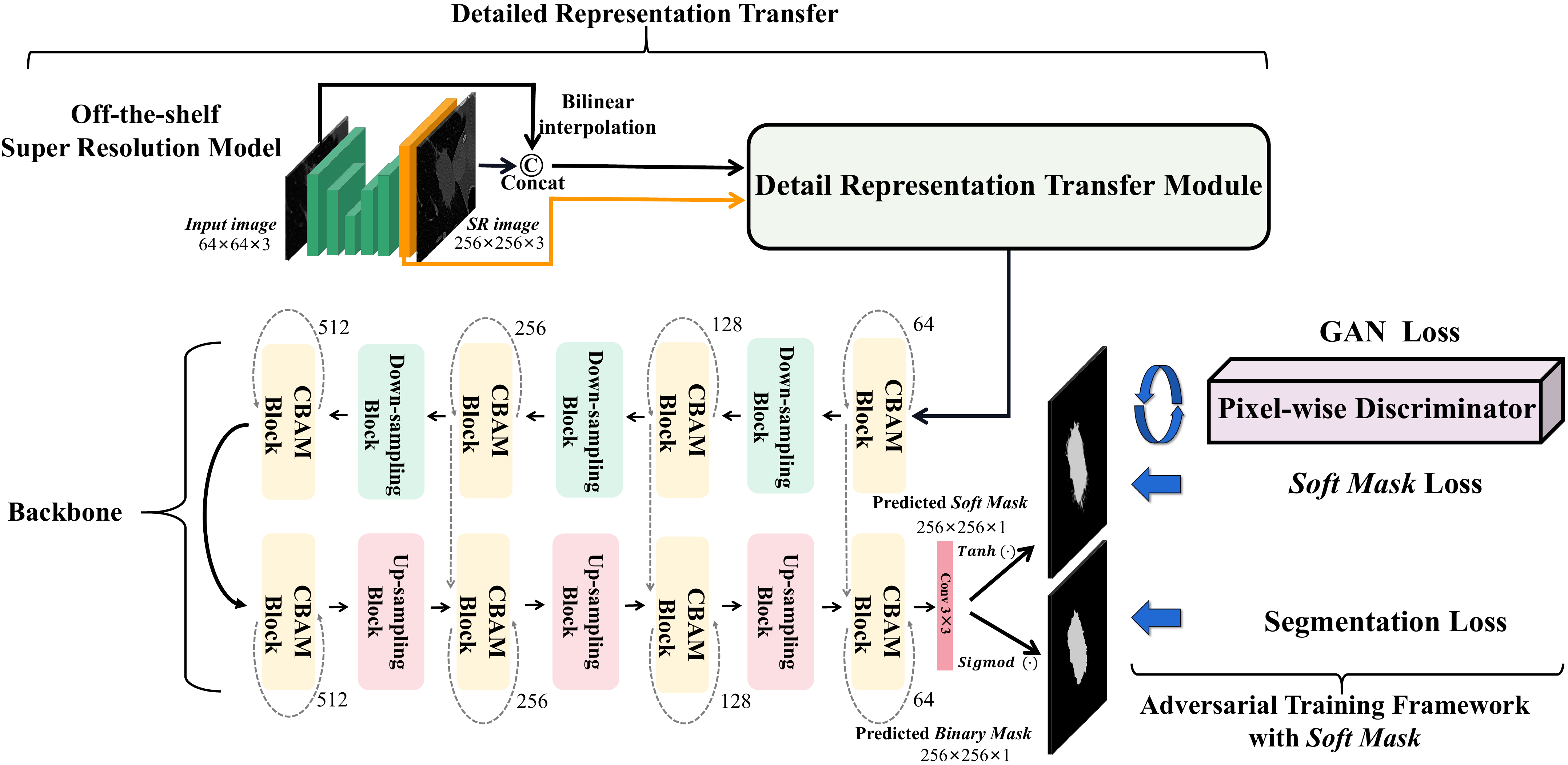}
\end{center}
   \caption{Overview of the our DSNet. Our DSNet is composed of detail representation transfer module, backbone network and adversarial training framework with \textit{Soft Mask}. 
}
\label{fig:DSNet}
\end{figure*}

\noindent\textbf{(3) Trimap Generation with RECIST marks:} 
RECIST marks~\cite{eisenhauer2009new} are commonly found in current hospital picture and archiving systems (PACS) despite their coarseness. This means that converting these massive RECIST marks data into accurate masks will have great application value and potential. As shown in Fig.~\ref{fig:mask}~(a), RECIST marks have a long axis and a short axis to mark the diameters of the lesion. First, we use the regions marked on the long axis and the short axis as the prior information of the GrabCut~\cite{grabcut} to obtain the initial rough binary masks. Then we apply the morphological processing mentioned in \textbf{strategy (1)} to get the trimap.

After obtaining the reliable trimap, we use the close-form-matting algorithm~\cite{cfm} to generate \textit{Soft Masks} shown in Fig.~\ref{fig:mask}~(c).
Since solving $\alpha$ for each pixel in Eq.~\ref{eq: eq1} is an ill-posed problem, appropriate conditional assumptions need to be made.
The close-form-matting algorithm~\cite{cfm} assumes that the foreground $L$ and background $B$ are almost unchanged in a small window centered on any pixel, i.e. the foreground and background satisfy local smoothness. Base on this assumption, Eq.~\ref{eq: eq1} can be re-expressed as:
\begin{align}
\alpha_{i}\approx aM_{i}+b, ~\forall~i \in w
\end{align}
where $a=\frac{1}{L-B}$,$B=-\frac{B}{F-B}$, and $w$ is a small window in image. This approximate expression suggests finding $\alpha$,$a$,and $b$ that minimize the cost function as:
\begin{align}
J(\alpha,a,b)=\sum_{j\in M}(\sum_{i\in w_{j}}(\alpha_{i}-a_{j}M_{i}-b_{j})^{2}+\epsilon a_{j}^{2})
\label{eq:eq2}
\end{align}
where $w_{j}$ is a small window around pixel $j$, and the $\epsilon a_{j}^{2}$ is a regularization term on $a$ to obtain numerically stable and smooth solutions. For an image with $N$ pixels, there are $3N$ unknowns in Eq.~\ref{eq:eq2}. After a series of derivations and proofs~\cite{cfm}, the cost function can be simplified to:
\begin{align}
J(\alpha)= \alpha^{T}L\alpha,
\label{eq:eq3}
\end{align}
where $L$ is a $N\times N$ matrix, whose $(i,j)$ entry is:
\begin{align}
\sum_{k|(i,j)\in w_{k}}(\delta_{i,j}-\frac{1}{|w_{k}|}(1+\frac{1}{\frac{\epsilon}{|w_{k}|}+\sigma_{k}^{2}}(M_{i}-\mu_{k})(M_{j}-\mu_{k}))).
%\label{eq:eq4}
\end{align}
Here, $\delta_{i,j}$ is the Kronecker delta, $\mu_{k}$ and $\sigma_{k}^{2}$ are the mean and
variance of the intensities in the window $w_{k}$ around $k$, while $|w_{k}|$ is the number of pixels in this window.
In general, the close-form-matting algorithm constructs a Laplacian weight matrix by assuming the smoothness of the local small window and exploiting the correlation between each pixel in the local small window. 
Then we take the explicit $\alpha$ given by trimap as a constraint, and the optimal value of the cost function can be obtained by solving the minimum eigenvector of the $L$ weight matrix, so as to obtain the solution of the entire $\alpha$ matrix. Finally, substitute $\alpha$ into Eq.~\ref{eq: eq1} to obtain the \textit{Soft Mask}.
%Then the optimal value of the cost function can be obtained by solving the minimum eigenvector of the weight matrix, and this minimum eigenvector is the prerequisite to calculate the \textit{Soft Mask} based on $\alpha$ matrix~\cite{cfm}.
%Then the optimal value of the cost function can be obtained by solving the minimum eigenvector of the weight matrix, and this minimum eigenvector is the $\alpha$ matrix required to calculate the \textit{Soft Mask}.
%We refer the readers to ~\cite{cfm} for more details. 

Based on the above automated pipeline, we label the lung nodules in the DeepLesion datasets~\cite{Yan:2018:DeepLesion} with RECIST marks to get \textit{Soft Masks}.
Then, we further binarize (with $0.5$ as the threshold) \textit{Soft Masks} to get an accurate binary mask (Fig.~\ref{fig:mask}~(d)) for the lung nodules segmentation task.
Besides, the entire pipeline is efficient, which only takes 20 $ms$ (for a single 256 $\times$ 256 image: trimap 5 $ms$, close-form-matting 15 $ms$) for us to obtain a high-quality \textit{Soft Mask} label. 
We label the lung nodules in the DeepLesion datasets~\cite{Yan:2018:DeepLesion} with RECIST marks and form a new dataset containing $1500$ lung nodules named \textit{Soft Mask} dataset of Lung Nodules (\textit{LNSM}). 

\subsection{Lung Nodule Segmentation Network with Detailed Representation Transfer and \textit{Soft Mask} Supervision (DSNet)}
Our DSNet can transform low-quality lung nodule images into a high-quality mask for accurate segmentation shown in Fig.~\ref{fig:DSNet}.
Specifically, we first design a Detailed Representation Transfer Module (DRTM) to collect feature maps with detailed representations as the input to the network. Then we employ a special backbone network to extract features and output prediction results.

\subsubsection{Backbone Architecture}
\begin{figure}[H]
\begin{center}
  \includegraphics[width=1 \linewidth]{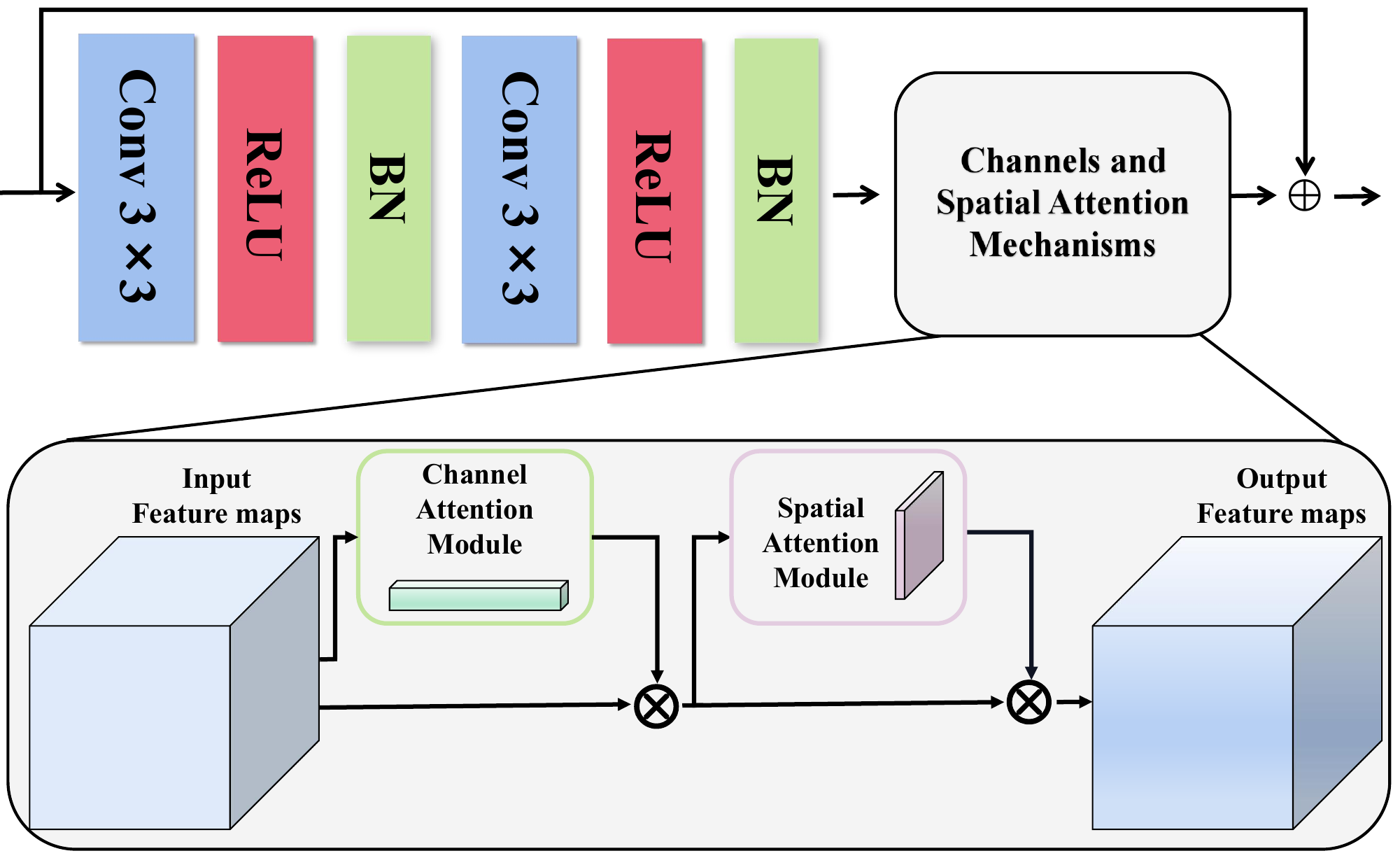}
\end{center}
   \caption{Details of the CBAM Block in Fig.~\ref{fig:DSNet}. Each $3\times3$ convolutional layer is followed by a ReLU activation function and a batch normalization layer.}
\label{fig:cbam}
\end{figure}

Based on the classic encoder-decoder architecture of U-Net~\cite{unet}, we design a special backbone to adapt to the accurate segmentation of lung nodules. 
As shown in Fig.~\ref{fig:DSNet}, our backbone consists of basic blocks, up-sampling blocks, and down-sampling blocks. The CBAM block~\cite{cbam} is employed as the basic block of our backbone shown in Fig.~\ref{fig:cbam}, which uses both channel attention and spatial attention to enhance the expressive ability of feature maps and thus stimulate the performance of segmentation. In addition, the shortcut connection of ResNet~\cite{resnet} is also utilized to ensure the propagation of detailed information and gradients.  In order to reduce the loss of detail caused by down-sampling on the input image whose resolution is usually low, we use $4 \times 4$ convolution layer with $stride = 2$, $padding = 1$ and ReLU activation function as the down-sampling block instead of pooling. The up-sampling 
block of the decoder consists of $4 \times 4$ transposed convolution and ReLU activation function, while the skip connection of U-Net is still retained to recover the detailed information.

\subsubsection{Segmentation Head} After a $3\times3$ convolutional layer, the predicted segmentation mask is generated and accepts the joint supervision of the binary mask (after \textit{Sigmod}) and the \textit{Soft Mask} (after \textit{Tanh}  which normalizes the result to [-1,1]). 
We also normalize the ground truth of \textit{Soft Mask} to the same range via the normalization function in Pytorch~\cite{paszke2019pytorch}.

\begin{figure*}[t]
\begin{center}
  \includegraphics[width=1 \linewidth]{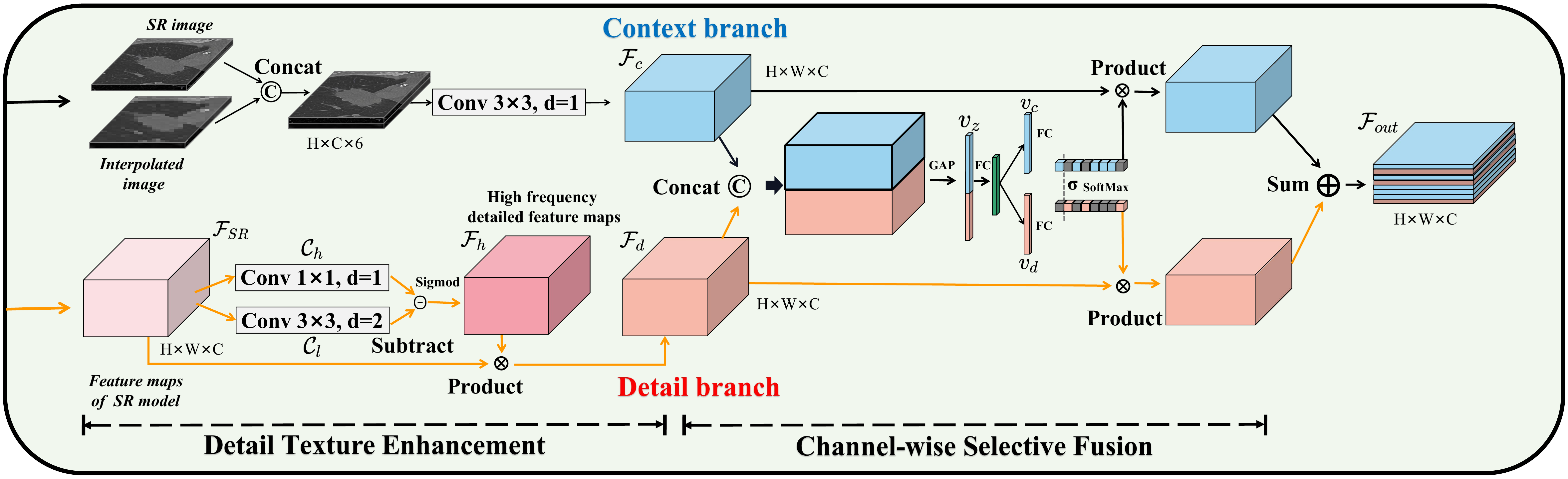}
\end{center}
   \caption{Detailed design of Detailed Representation Transfer Module (DRTM). Our DRTM includes two operations, detail texture enhancement and channel-selective fusion, to adaptively capture detailed representations from super-resolution (SR) models.
}
\label{fig:drtm}
\end{figure*}
%\vspace{-0.2cm}

\subsubsection{Detailed Representation Transfer}
\label{sec: drt}
Lung nodules are usually very small (5 $ \sim $ 10 mm), so common CT systems can only obtain images of lung nodules with low resolution. 
The low-resolution image means the loss of detailed information, which is challenging for accurate segmentation of lung nodules with complex shapes and blurred edges.
Inspired by super-resolution methods,
we reconstruct the detailed information in the high-resolution image and use it to stimulate the performance of accurate segmentation. Next we will show how to train a super-resolution model without high-resolution lung nodules data and apply the detailed information from the off-the-shelf super-resolution model to the segmentation pipeline. 

\noindent \textbf{(1) Self-supervised super-resolution model:}  
The lack of high-resolution lung nodules images makes it difficult to train a super-resolution model. Due to the robustness of the super-resolution methods~\cite{srgan} and the inspiration of the self-supervised pre-training methods~\cite{moco}, we self-supervised train a super-resolution module with a general CT images dataset DeepLesion~\cite{Yan:2018:DeepLesion}. The DeepLesion dataset consists of $32,735$ important clinical radiological findings (lesions, tumors, lymph nodes, etc.), and it can be universally adapted to the data of different organs. Specifically, we down-sample the CT images of the Deeplesion dataset from  $512 \times 512$ to $128 \times 128$ for training an SRGAN model~\cite{srgan} (from 128 to 512). Then the trained off-the-shelf super-resolution model is used as a teacher model to transfer the learned detailed knowledge to the Detailed Representation Transfer Module (DRTM) for further lung nodules segmentation.

For transferring the detailed representation into the segmentation pipeline, we designed our DRTM to capture detailed representations as follows.

\noindent \textbf{(2) Detailed Representation Transfer Module:} 
First, our DRTM uses the context branch and the detail branch to extract feature maps respectively, as shown in Fig.~\ref{fig:drtm}.
For the context branch, we combine the interpolated low-resolution image and the output of the SR model and use the convolutional layer to extract the feature maps $\mathcal{F}_{c}$.
For the detail branch, we explicitly mine detail representations from the feature map output of the last block of the SR model decoder.
Specifically, given the input feature maps from SR model
$\mathcal{F}_{SR} \in \mathbb{R}^{H\times W \times C}$, we use two groups of convolutions to extract features from different receptive fields. 
We explicitly model high-frequency detail information by subtracting the outputs of these two convolutional layers as follows:
\begin{align}
\mathcal{F}_{h}=sigmoid(C_{h}(\mathcal{F}_{SR})-C_{l}(\mathcal{F}_{SR})),
%\label{eq:eq4}
\end{align}
where $C_{h}$ denotes $1\times 1$ point-wise convolution and $C_{l}$ denotes $3\times 3$  dilated convolution with dilated rate $2$.
Here we use point-wise convolution $C_{h}$ to preserve local high-frequency details, and employ dilated convolution $C_{l}$ with a large receptive field to extract low-frequency semantic structure information.
Therefore, the high-contrast pixels obtained by their subtraction are regarded as high-frequency detail information. Then $\mathcal{F}_{h}$ is utilized to weight the original feature map to achieve the detailed texture enhancement as:
\begin{align}
\mathcal{F}_{d}=\mathcal{F}_{SR}\times\mathcal{F}_{h},
\end{align}
where $\mathcal{F}_{d}$ is the feature maps output by the detail branch. 

\begin{figure}[H]
\begin{center}
  \includegraphics[width=1 \linewidth]{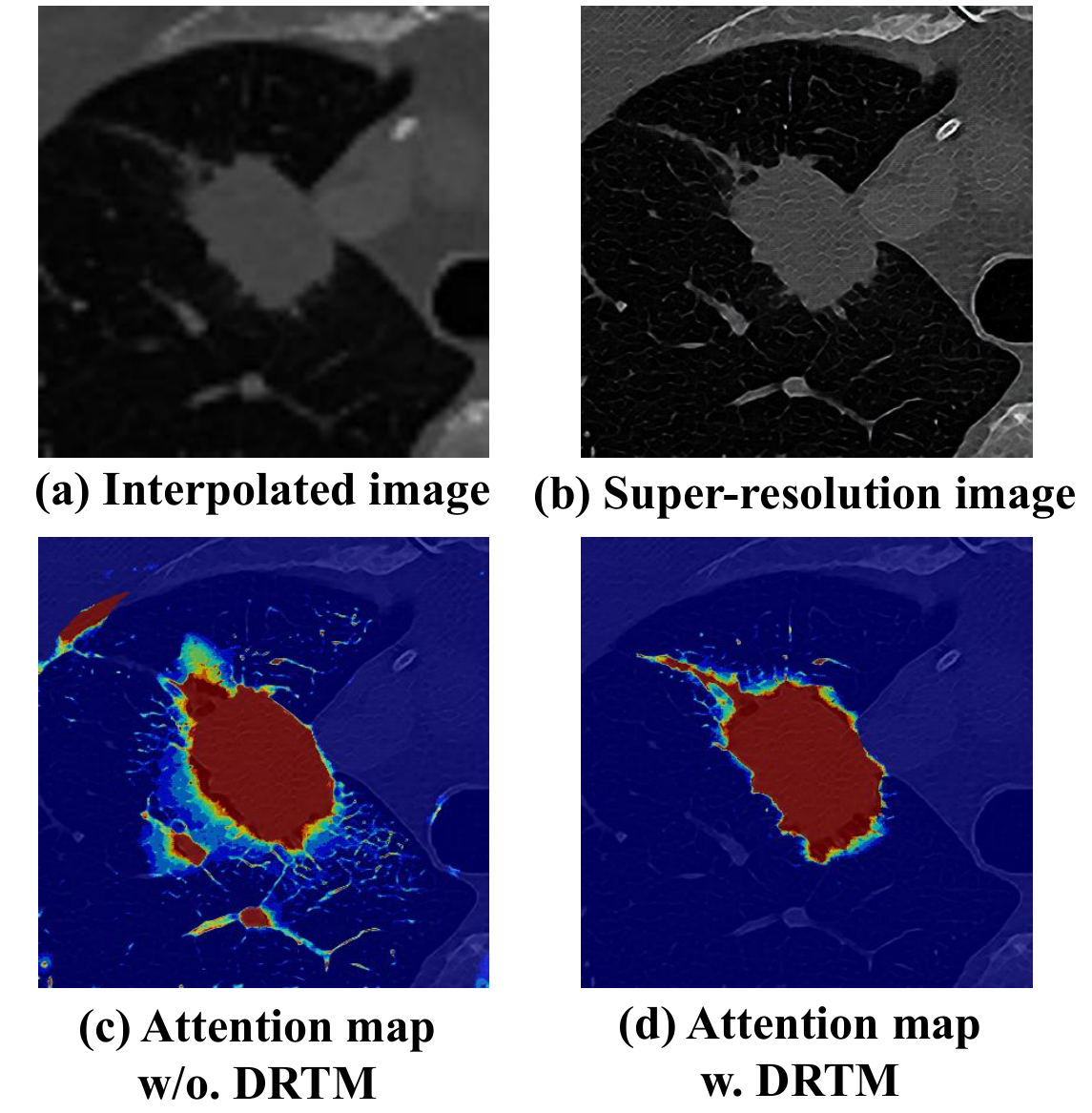}
\end{center}
   \caption{Some visualization samples. DRTM: Detailed Representation Transfer Module.}
\label{fig:samples}
\end{figure}

However, not all detailed representations from the super-resolution model are conducive to the segmentation, and some noises also are introduced shown in Fig.~\ref{fig:samples} (b).
To suppress the noise and better fusion the detailed representation, we further propose a novel channel-wise selective fusion operation. 
Specifically, two compact vectors $v_{c}, v_{d} \in \mathbb{R}^{1\times C}$ are created to enable the guidance for the precise and adaptive selections. 
This is achieved by the simple Global Average Pooling (GAP) and fully connected (FC) layer:
\begin{align}
v_{z}=GAP(Concat(\mathcal{F}_{d},\mathcal{F}_{c})),
\end{align}
where $Concat$ denotes merging two feature maps along the channels.
Then we use GAP to generate vector $v_{z} \in \mathbb{R}^{1 \times 2C}$.
\begin{align}
v_{c},v_{d}=FC(v_{z}),
\end{align}
Here the fully connected layer (FC) reduces $v_{z}$ to $\mathbb{R}^{1 \times C}$ dimension and then predicts two weight vectors $v_{c}$, $v_{d}$ for the two branches respectively.

Then, a soft attention across channels is used to select different spatial scales of information adaptively, which is
guided by the compact vectors $v_{c},v_{d}$. Specifically, a
softmax operator is applied on the channel-wise digits:
\begin{align}
a = \frac{e^{v_{c}}}{e^{v_{c}}+e^{v_{d}}},b = \frac{e^{v_{d}}}{e^{v_{c}}+e^{v_{d}}}
\end{align}
where $a,b \in \mathbb{R}^{1 \times C}$ denote the soft attention vectors for $\mathcal{F}_{c}$ and $\mathcal{F}_{d}$, respectively.
The final output feature maps $\mathcal{F}_{out} \in \mathbb{R}^{H \times W \times C}$ is obtained through the weighted fusion:
\begin{align}
\mathcal{F}_{out}= a \cdot \mathcal{F}_{c} + b \cdot \mathcal{F}_{d},~ a+b=1
\end{align}

In addition, Fig.~\ref{fig:samples}~(c) and (d) show the spatial attention maps in the first CBAM block of DSNet.
Here sample (c) is to directly use the feature maps $\mathcal{F}_{SR}$ of the super-resolution model as input while sample (d) is to use DRTM for representation transfer. 
We can see more impressive edge details (thanks to detail texture enhancement) and less sharpening noises (thanks to channel-wise selective fusion) with the help of our DRTM.

\subsection{Adversarial training framework with \textit{Soft Mask}.}

\noindent We use DICE loss~\cite{vnet} as the \textbf{Segmentation Loss} for the supervision of ground truth binary masks:

\begin{align}
\mathcal L_{\rm Seg}=1-\frac{2\sum_{i=1}^{N}(x_{i}{y}_{i})}{\sum_{i=1}^{N}x_{i}^{2}+\sum_{i=1}^{N}{y}_{i}^{2}}.
\label{equ:segloss}
\end{align}
where $y_{i}$ is the ground truth segmentation mask $Y_{bin}$ for a given pixel $i$, and ${x}_{i}$ is the corresponding value in the predicted binary mask $X_{bin}$.

\textbf{\textit{Soft Mask} Loss}  is given to make the predicted \textit{Soft Mask} has the same distribution as the ground truth \textit{Soft Mask}:
\begin{align}
\mathcal L_{\rm Soft}=||X_{soft}-Y_{soft}||_{L1}
\label{equ:softloss}
 \end{align}\\ 
Where $X_{soft}$ and $Y_{soft}$ denote ground truth \textit{Soft Mask} and predicted \textit{Soft Mask} respectively, and $||\cdot||_{L1}$ denotes the $L1$ distance.
\begin{figure}[H]
\begin{center}
  \includegraphics[width=1 \linewidth]{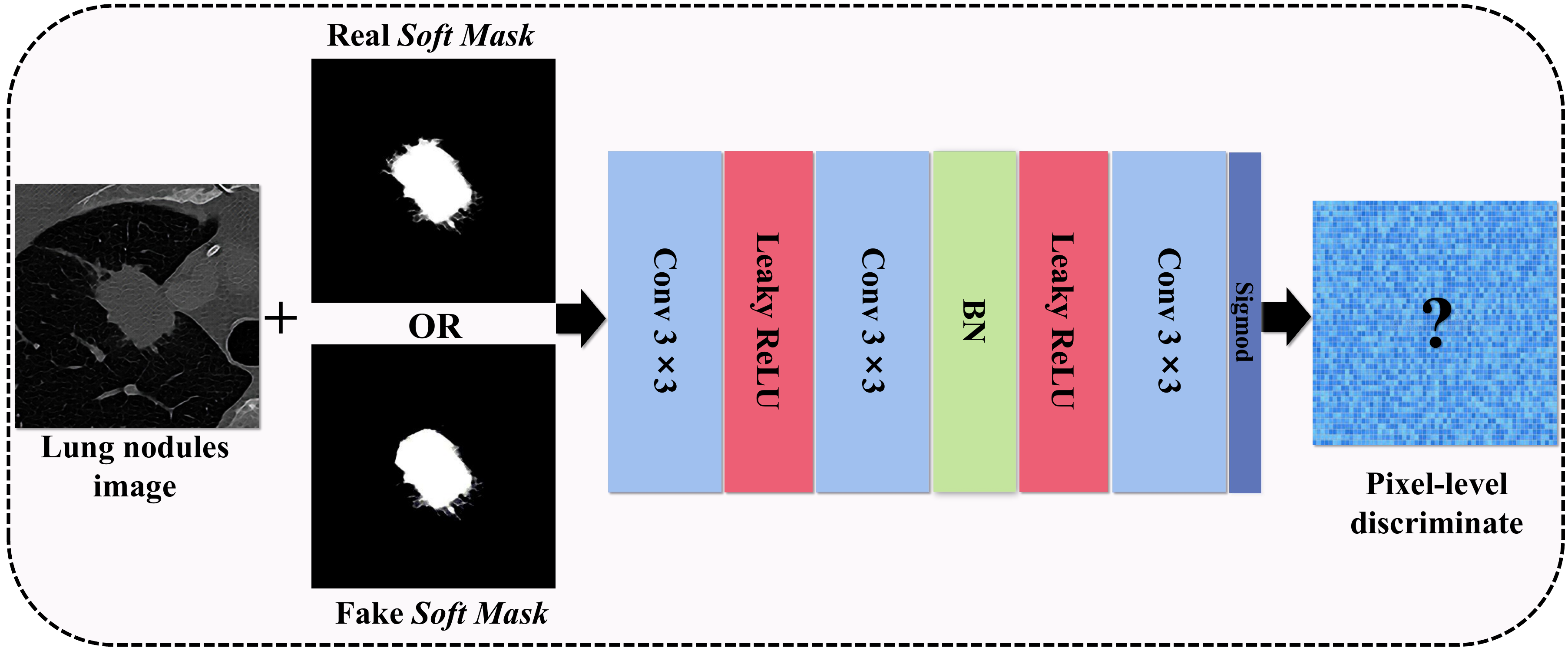}
\end{center}
   \caption{The structure of the pixel-wise Discriminator.}
\label{fig:netd}
\end{figure}
\textbf{GAN Loss} is introduced to further improve the quality of the predicted \textit{Soft Mask} by conditional generative adversarial supervision~\cite{cgan}. As shown in Fig~\ref{fig:netd}, a pixel-wise Discriminator is designed to provide adversarial GAN loss for DSNet. Notice that pixel-level discrimination is utilized to obtain an accurate pixel-wise adversarial loss.
Specifically, the overall objective can be expressed as:
\begin{align}
G^{*}=arg\,\underset{S}{min}\,\underset{D}{max}\,\mathcal L_{GAN}(S,D),
\end{align}
\begin{align}
\mathcal L_{\rm GAN} = logD(x,y)+log(1-D(x,S(x))
\end{align}
where our DSNet \textit{S} tries to minimize this objective against an adversarial Discriminator \textit{D} that tries to maximize it. The \textit{x} denotes the lung nodules image (condition), \textit{y} denotes the real \textit{Soft Mask} (ground truth), and \textit{S(x)} denotes the fake \textit{Soft Mask} (predicted \textit{Soft Mask}).
In the training framework, our DSNet and the Discriminator have trained alternately.

\subsection{Implementation}
The total loss function of our DSNet can be defined as :
\begin{align}
\mathcal L_{total} = \lambda_{1} \mathcal L_{\rm Seg} + \lambda_{2} \mathcal L_{\rm Soft} + \lambda_{3} \mathcal L_{\rm GAN}
\end{align}
where $\lambda_{1}$,$\lambda_{2}$,$\lambda_{3}$ are empirically set to $0.5$, $100$, $1$.
The Adam optimizer~\cite{kingma2014adam} with poly learning rate policy is used to optimize the network with the training batch size set to $4$. The learning rate decays from $2e^{-4}$,
and the whole training process for all datasets  typically converges in about 100 epochs with a single NVIDIA Titan V GPU.

% Experiment
\section{Experiments}

\subsection{Accurate segmentation for lung nodules}
In this section, we have conducted comprehensive experiments on the proposed \textit{LNSM} and LIDC~\cite{lidc} datasets to verify the effectiveness of our method on lung nodules segmentation.
Following previous works~\cite{cfnet} and ~\cite{dual},  DICE similarity (DICE), sensitivity (SEN) and positive predictive value (PPV) are used as the evaluation metrics. 
Here DICE score is a region-level similarity measure that mainly focuses on the internal structural consistency of segmented objects, while SEN and PPV are pixel-level evaluation measures that equally consider the influence of each pixel to give the error value between the predicted mask and the ground truth.
For convenient, we use the following denotations: $TP$ (true positive), $TN$ (true negative), $FP$ ( false positive), and $FN$ (false negative).
Therefore, these metrics are correspondingly defined as: 
\begin{align}
DICE =\frac{2\times TP}{2\times TP+FP+FN},
\end{align}
\begin{align}
SEN= \frac{TP}{TP+FN},
\end{align}
\begin{align}
PPV= \frac{TP}{TP+FP}.
\end{align}
\begin{figure}[t]
\begin{center}
  \includegraphics[width=1 \linewidth]{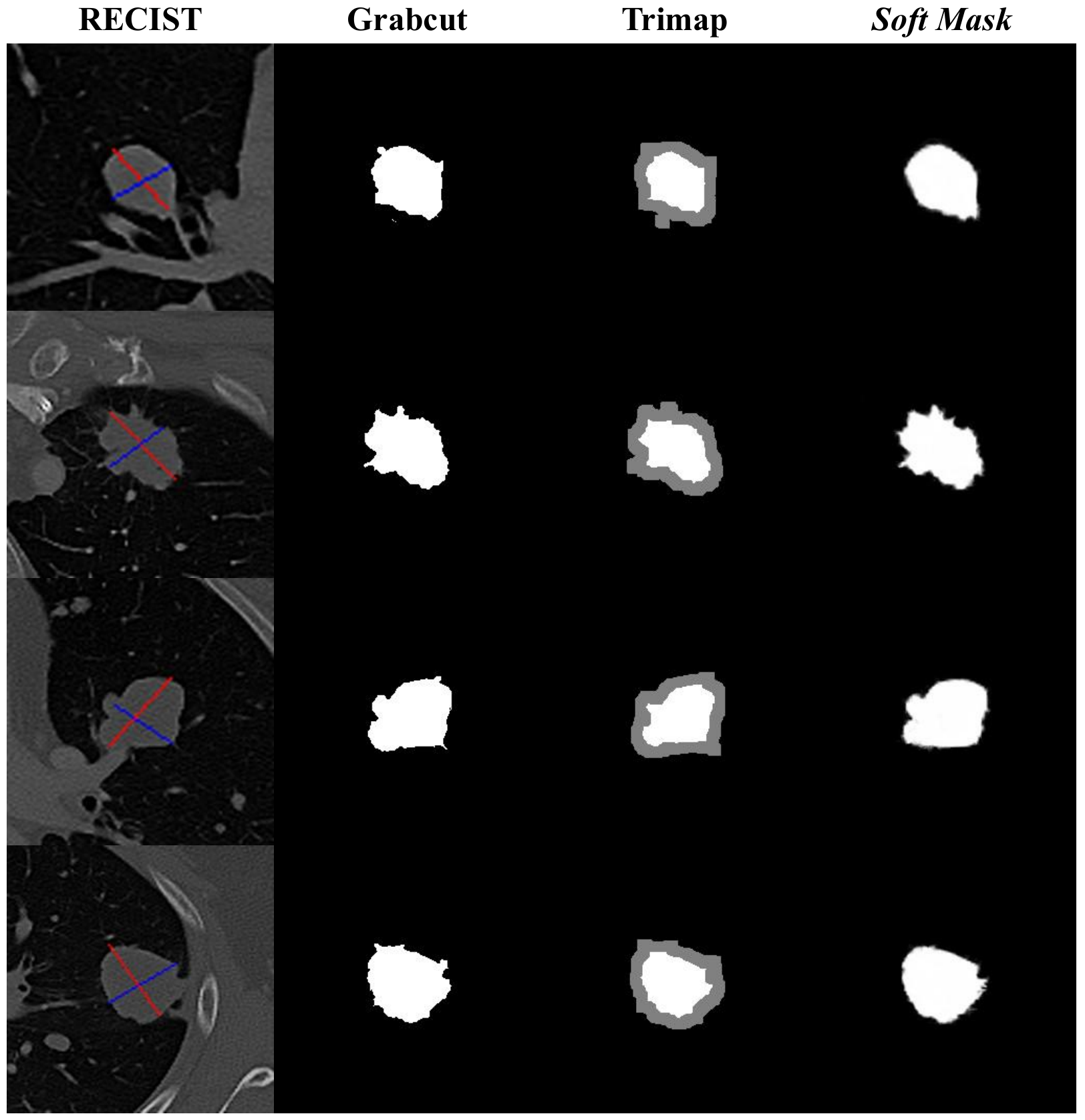}
\end{center}
   \caption{Examples of \textit{Soft Mask} annotation pipeline of LNSM dataset. }
\label{fig:lnsm}
\end{figure}

\subsubsection{Comparison on our \textit{LNSM} dataset}
Lung nodule images with RECIST marks in Deeplesion dataset~\cite{Yan:2018:DeepLesion} are used to build our \textit{LNSM} dataset. According to the method (\textbf{strategy (3)} in Sec.~\ref{sec: define}) proposed in the previous section, we obtained $1500$ cropped $64\times64$ lung nodules images with \textit{Soft Masks} and binary masks for training supervision.
We show in Fig.~\ref{fig:lnsm} the steps to automatically obtain the \textit{Soft Mask} via the RECIST marks.
We randomly divide the dataset into training sets (1,000 nodules), test sets (400 nodules), and validation sets (100 nodules), while all methods use the same dataset division.
In the test, all methods utilize the same binary mask in  \textit{LNSM} as the ground truth to calculate metrics for a fair comparison.

\begin{table}[thb]\small
\centering
%\vspace{-0.4cm}
\caption{Evaluation results on our \textit{LNSM} dataset. }
\setlength{\tabcolsep}{1.5mm}{
\renewcommand\arraystretch{1.5} {
\begin{tabular}{llll} 
\hline
Methods       & DICE (\%)                 & SEN (\%)                  & PPV (\%)                   \\ 
\hline
UNet~\cite{unet}          & 87.88 $\pm $ 1.1          & 86.62 $\pm $ 3.2          & 89.04 $\pm $ 5.7           \\
CF-CNN~\cite{cfnet}        & 88.98~$\pm $ 2.5          & 87.20 $\pm $ 3.2          & 89.60 $\pm $ 2.3           \\
DB-ResNet~\cite{dual}  & 89.21 $\pm $ 2.2          & 88.22 $\pm $ 1.6          & 90.04 $\pm $~1.8           \\
UNet++~\cite{unet++}        & 88.57 $\pm $ 2.7          & 89.24 $\pm $ 0.5          & 91.86 $\pm $ 3.7           \\
INFNet~\cite{infnet}        & 90.05 $\pm $ 1.8          & 88.17 $\pm $ 2.3          & 93.41 $\pm $ 3.7           \\
INFNet$^{\dag}$~\cite{infnet}        & 90.53 $\pm $ 1.7          & 89.21 $\pm $ 1.3          & 94.04 $\pm $ 2.5           \\
MedT~\cite{medt}          & 90.13 $\pm $ 2.6          & 89.56 $\pm $ 3.1          & 92.41 $\pm $ 1.2           \\ 
MedT$^{\dag}$~\cite{medt}          & 91.02 $\pm $ 2.4          & 90.10 $\pm $ 1.5          & 93.12 $\pm $ 2.7           \\ 
\hdashline
DSNet (our) & \textbf{93.77 $\pm $ 3.1} & \textbf{91.76 $\pm $ 0.9} & \textbf{95.88 $\pm $ 2.5}  \\
\hline
\end{tabular}}}
%\vspace{-0.2cm}
\label{tab:lnsm}
\end{table}

Table~\ref{tab:lnsm} presents a quantitative comparison of some advanced methods on \textit{LNSM} dataset with the same dataset settings. The outputs are in “ mean $\pm$ standard deviation"
format. Specifically, UNet~\cite{unet}, CF-CNN~\cite{cfnet}, DB-ResNet~\cite{dual}, UNet++~\cite{unet++}, INFNet~\cite{infnet}, and recent MedT~\cite{medt} are compared.
Here $\dag$ represents using SR image output by super-resolution model we got in Sec.~\ref{sec: drt} (1) as the input.
We can see that our DSNet exceeds these state-of-the-art methods with
a large margin, especially on the DICE and PPV metric scores.

\subsubsection{Comparison on LIDC}
 We used a public lung nodules CT dataset from the Lung Image Database Consortium and Image Database Resource Initiative (LIDC)~\cite{lidc} for further comparison. In this study, we studied $986$ nodule samples annotated by four radiologists. Due to the differences in labeling between the four radiologists, the 50 \% consensus criterion~\cite{2011Segmentation} was used to generate the ground truth binary masks. We use the method ({\textbf{strategy(2) in Sec.~\ref{sec: define}}}) of labeling \textit{Soft Mask} to obtain ground truth \textit{Soft Masks} for training supervision.
Then, we randomly partition these nodules into three subsets for training, validation, and testing with the number of nodules contained in each subset being $387$, $55$, and $544$, respectively. 
 
\begin{figure*}[t]
\begin{center}
  \includegraphics[width=0.9 \linewidth]{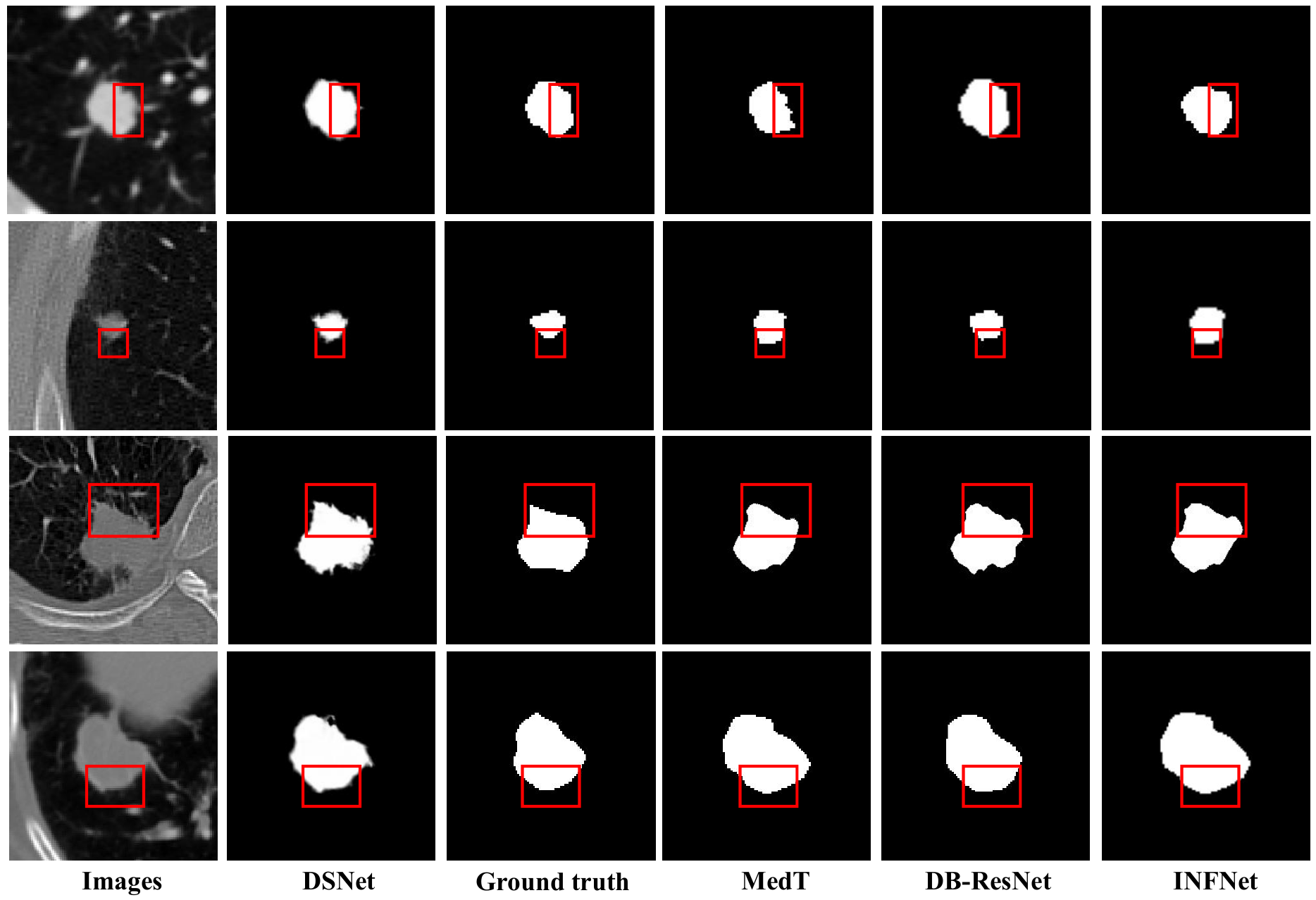}
\end{center}
   \caption{Visual comparison between our DSNet and advanced methods on LIDC~\cite{lidc} dataset.
   Thanks to the \textit{Soft Mask} supervision, {our DSNet has an impressive visual effects
that even exceeds binary ground truth.}
}
\label{fig:lidc_vis}
\end{figure*}

We compared our DSNet with some advanced segmentation methods including UNet~\cite{unet}, CF-CNN~\cite{cfnet}, DB-ResNet~\cite{dual}, UNet++~\cite{unet++}, INFNet~\cite{infnet}, and MedT~\cite{medt}, which are illustrated in Tab.~\ref{tab:lidc}. 
Noted that $\dag$ represents using SR image output by super-resolution model we got in Sec.~\ref{sec: drt} (1) as the input.
The results show that our method outperforms other methods in all metrics.
Compared with the most competitive MedT$^{\dag}$,
our method leads  2.81\%, 2.23\%, and 3.68\% respectively.
Besides,  the visual comparison between our DSNet and these methods on the LIDC dataset is shown in Fig.~\ref{fig:lidc_vis}.
\begin{table}[thb]\small
\centering
\caption{Evaluation results on the LIDC dataset.}
\setlength{\tabcolsep}{1.5mm}{
\renewcommand\arraystretch{1.5} {
\begin{tabular}{llll} 
\hline
Methods       & DICE (\%)                  & SEN (\%)                   & PPV (\%)                    \\ 
\hline
UNet~\cite{unet}          & 77.84 $\pm$ 21.7          & 77.98 $\pm$ 24.5          & 82.52 $\pm$ 21.5           \\
CF-CNN~\cite{cfnet}        & 78.55 $\pm$ 12.5          & 86.01 $\pm$ 15.2          & 75.79 $\pm$ 14.7           \\
DB-ResNet~\cite{dual}     & 82.74 $\pm$ 10.2           & 89.05 $\pm$ 11.8          & 79.64 $\pm$ 13.5           \\
UNet++~\cite{unet++}        & 80.54 $\pm$ 12.9          & 87.96 $\pm$ 17.2          & 79.18 $\pm$ 15.1           \\
INFNet~\cite{infnet}        & 81.01 $\pm$ 14.1          & 88.33 $\pm$ 13.7          & 75.58 $\pm$ 17.6           \\
INFNet$^{\dag}$~\cite{infnet}        & 81.67 $\pm$ 13.7          & 87.25 $\pm$ 10.5          & 76.24 $\pm$ 13.3           \\
MedT~\cite{medt}          & 81.34 $\pm$ 7.9          & 87.96 $\pm$ 11.4          & 80.14 $\pm$ 15.3           \\ 
MedT$^{\dag}$~\cite{medt}          & 82.08 $\pm$ 6.5          & 88.33 $\pm$ 11.8          & 81.01 $\pm$ 16.4           \\ 
\hdashline
DSNet (our)& \textbf{84.89 $\pm $ 7.2} & \textbf{90.56 $\pm$ 12.0} & \textbf{84.69 $\pm$ 13.4}  \\
\hline
\end{tabular}}

}
\label{tab:lidc}
\end{table}
%\vspace{-0.3cm}

\subsection{Ablation study}
To validate the effectiveness of our modules and strategies, we conduct the following ablation studies on both \textit{LNSM} and LIDC datasets.

\begin{table}[thb]\small
\caption{Ablation study on the \textit{LNSM} and LIDC benchmarks. We report the \emph{DICE} scores for different variants of our DSNet.
%on the Context Augmented and Spatial Attention Weighted.
% \textbf{SR Input} means using the lung nodules images processed by the super-resolution model as input.
% \textbf{DRTM} means Detailed  Representation  Transfer  Module.
}
\setlength{\tabcolsep}{1.3mm}{
\renewcommand\arraystretch{1.5}     
{
\begin{tabular}{c|c|c|c} 
\hline
\multicolumn{2}{c|}{Method}                                                                                                                                                       & \textit{LNSM}              & LIDC                          \\ 
\hline
\textit{Config.}                                                                                            & \textit{Variants}                                                   & \textit{DICE} (\%)         & \textit{DICE} (\%)            \\ 
\hline
\multirow{2}{*}{\begin{tabular}[c]{@{}c@{}}Backbone\\ (Baseline)\end{tabular}}                              & \textit{orig. UNet}                                                 & 87.88 $\pm$ 1.1            & 77.84 $\pm$ 21.7             \\ 
\cdashline{2-4}
                                                                                                            & \begin{tabular}[c]{@{}c@{}}our proposed\\ backbone\end{tabular}     & \textbf{89.57 $\pm$ 1.3 }  & \textbf{ 79.3 $\pm$ 25.3 }  \\ 
\hline
\multirow{2}{*}{\begin{tabular}[c]{@{}c@{}}+ Representation \\Transfer\end{tabular}}             & + SR Input                                                          & 90.15 $\pm$ 2.1            & 81.69 $\pm$ 15.7             \\
                                                                                                            & + DRTM                                                              & \textbf{ 91.83 $\pm$ 1.7 } & \textbf{83.88 $\pm$ 8.5 }    \\ 
\hline
\multirow{2}{*}{\begin{tabular}[c]{@{}c@{}}+ Adversarial\\ training with\\ \textit{Soft Mask}\end{tabular}} & \begin{tabular}[c]{@{}c@{}}+ \textit{Soft Mask}\\ Loss\end{tabular} & 93.26 $\pm$ 2.0            & 84.21 $\pm$ 10.3             \\ 
                                                                                                            & \begin{tabular}[c]{@{}c@{}}+ GAN Loss\\ (our DSNet)\end{tabular}    & \textbf{93.77 $\pm$ 3.1}   & \textbf{84.89 $\pm$ 7.2}     \\
\hline
\end{tabular}}}

\label{tab:abl}
\end{table}

\subsubsection{Comprehensive Ablation Study}

The ablation study is reported in Tab.~\ref{tab:abl}. 
We use the classic UNet~\cite{unet} trained with DICE loss as the baseline. After replacing with our backbone, the performance is significantly improved, which implies that the network structure is still crucial to lung nodules segmentation. Just using super-resolution images as the input, the DICE score has also been slightly improved, indicating that high-resolution images can promote the accurate segmentation of lung nodules. Note that our DRTM further improves the performance indicating the importance of detail texture enhancement and channel-wise selective fusion. Finally, our DSNet achieves the best performance after applying complete
adversarial training with \textit{Soft Mask} supervision.

\begin{table}[H]\small
\caption{Comparison of different soft labels.
}
\centering
\setlength{\tabcolsep}{1.5mm}{
\renewcommand\arraystretch{1.3}     {
\begin{tabular}{lll} 
\hline
Soft Labels                                     & LNSM~DICE(\%)              & LIDC~DICE(\%)               \\ 
\hline
w/o. soft label supervision                       & 91.23 $\pm$ 1.7          & 82.88 $\pm$ 8.5          \\
Soft label in~\cite{softlabel} & 91.92 $\pm$ 2.5          & 83.11 $\pm$ 8.6           \\
Soft label in~\cite{softseg} & 92.66 $\pm$ 1.4          & 83.95 $\pm$ 10.3           \\ 
\hdashline
Our \textit{Soft Mask}                                   & \textbf{93.77 $\pm $ 3.1} & \textbf{84.89 $\pm$ 7.2}  \\
\hline
\end{tabular}}}

\label{tab:softlabel}
\end{table}

\subsubsection{Ablation Study on Soft Mask Supervision}
In addition, Tab.~\ref{tab:softlabel} reports the comparison between our \textit{Soft Mask} and other soft labels~\cite{softlabel,softseg}.
Specifically, we use them to supervise the training of our DSNet under the same conditions, respectively.
The results demonstrate that our \textit{Soft Mask} is significantly better than other soft labels, due to the more accurate edge representation of our \textit{Soft Mask} which reduces the impact of imprecise boundary annotation.

\begin{table}
\centering
\caption{Ablation study on our DRTM. DTM: detail texture enhancement; CSF: channel-wise selective fusion.}
\label{tab:abl_DRTM}
\setlength{\tabcolsep}{3mm}{
\renewcommand\arraystretch{1.3}     {
\begin{tabular}{cccc} 
\hline
\multicolumn{2}{c}{Operations}                             & LNSM     & LIDC      \\ 
\hline
DTM & CSF & DICE(\%) & DICE(\%)  \\ 
\hline
                           &                               & 92.62 $\pm$ 1.5       & 83.64  $\pm$ 8.4       \\
\CheckmarkBold                          &                               & 93.41 $\pm$ 1.2       & 84.26 $\pm$ 7.5        \\
\CheckmarkBold                          & \CheckmarkBold                            & \textbf{93.77 $\pm$ 3.1}    & \textbf{84.89 $\pm$ 7.2}     \\
\hline
\end{tabular}}}
\end{table}

\subsubsection{Ablation Study on Detail Representation Transfer}
We also perform a meticulous ablation study for each component in our DRTM, as shown in Tab.~\ref{tab:abl_DRTM}.
The results show that the optimal DICE score is achieved when both operations detail texture enhancement and channel-wise selective fusion are applied.

\begin{figure*}[t]
\begin{center}
  \includegraphics[width=1\linewidth]{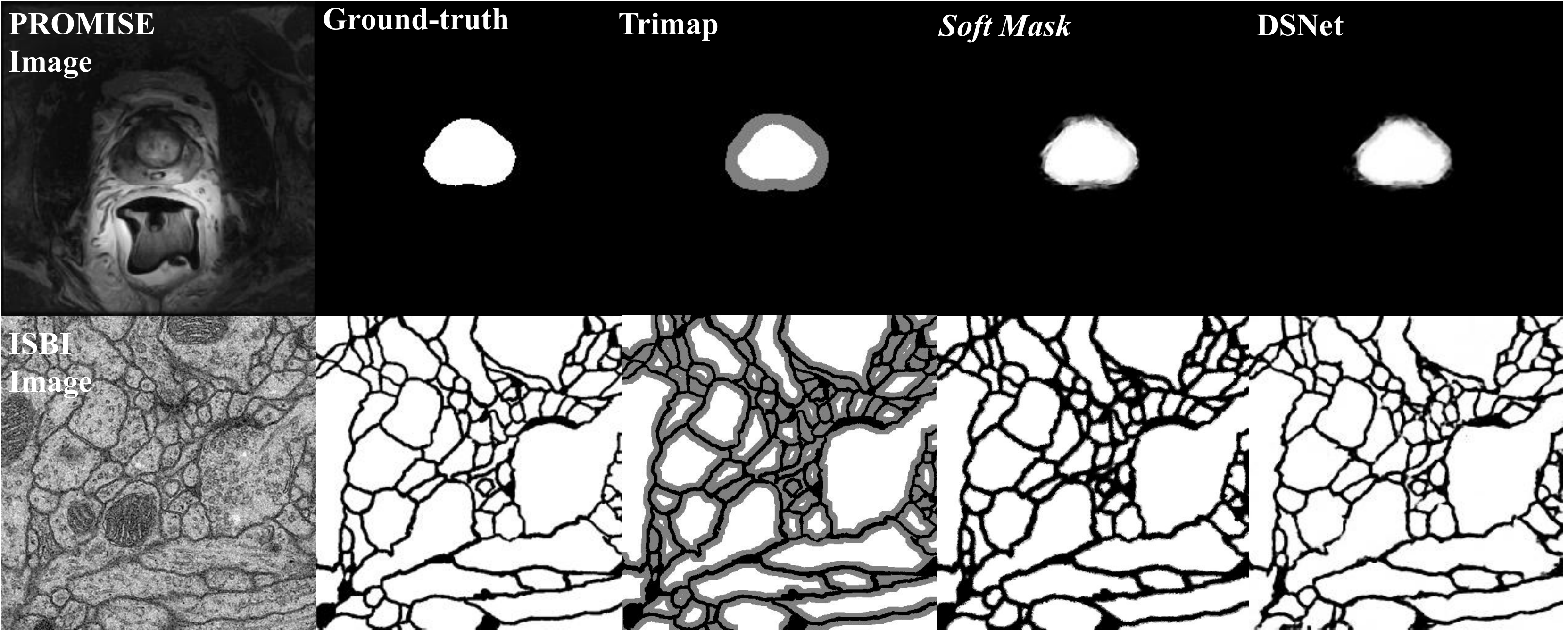}
\end{center}
   \caption{Some visual results on ISBI and PROMISE. From left to right are the original image, the ground truth binary mask of the dataset, the trimap image (Obtained by labeling strategy \textbf{(1)}), and the \textit{Soft Mask} and the results of our DSNet prediction.}
\label{fig:isb}
\end{figure*}
\subsection{Robustness Study}
\label{sec: expr}
To further verify the accurate segmentation performance and generalization of our method, we design an ambitious and challenging experiment: compare our DSNet with human doctors. Specifically, we compare our DSNet trained on \textit{LNSM} dataset with four radiologists on LIDC. Note that our DSNet is not retrained, which means that the training set and the test set are cross-domain.

\begin{table}[H]\small
\renewcommand\arraystretch{1}  
\centering
\caption{Compare our DSNet with four human doctors on LIDC dataset. D1, D2, D3, D4 mean Doctor 1, Doctor 2, Doctor 3, Doctor 4 respectively. 
}
\setlength{\tabcolsep}{1.5mm}{
\renewcommand\arraystretch{1.5} {
\begin{tabular}{c|cccccc} 
\hline
DICE (\%)                                                   & D1    & D2    & D3    & D4    & \begin{tabular}[c]{@{}c@{}}50\%\\~Consensus~\end{tabular} & DSNet           \\ 
\hline
D1                                                          & -     & 70.65 & 73.14 & 62.57 & 77.51                                                     & 77.76           \\
D2                                                          & 70.65 & -     & 74.46 & 64.91 & 82.92                                                     & 74.30           \\
D3                                                          & 73.14 & 74.46 & -     & 76.31 & 83.01                                                     & 80.88           \\
D4                                                          & 62.57 & 64.91 & 76.31 & -     & 69.74                                                     & 69.71           \\
\begin{tabular}[c]{@{}c@{}}50\% \\Consensus~ ~\end{tabular} & 77.51 & 82.92 & 83.01 & 69.74 & -                                                         & \textbf{86.35}  \\
DSNet                                                       & 77.76 & 74.30 & 80.88 & 69.71 & \textbf{86.35}                                            & -               \\
\hline
\end{tabular}
}}

\label{tab:robust}
\end{table}

\begin{figure}[H]
\begin{center}
  \includegraphics[width=1 \linewidth]{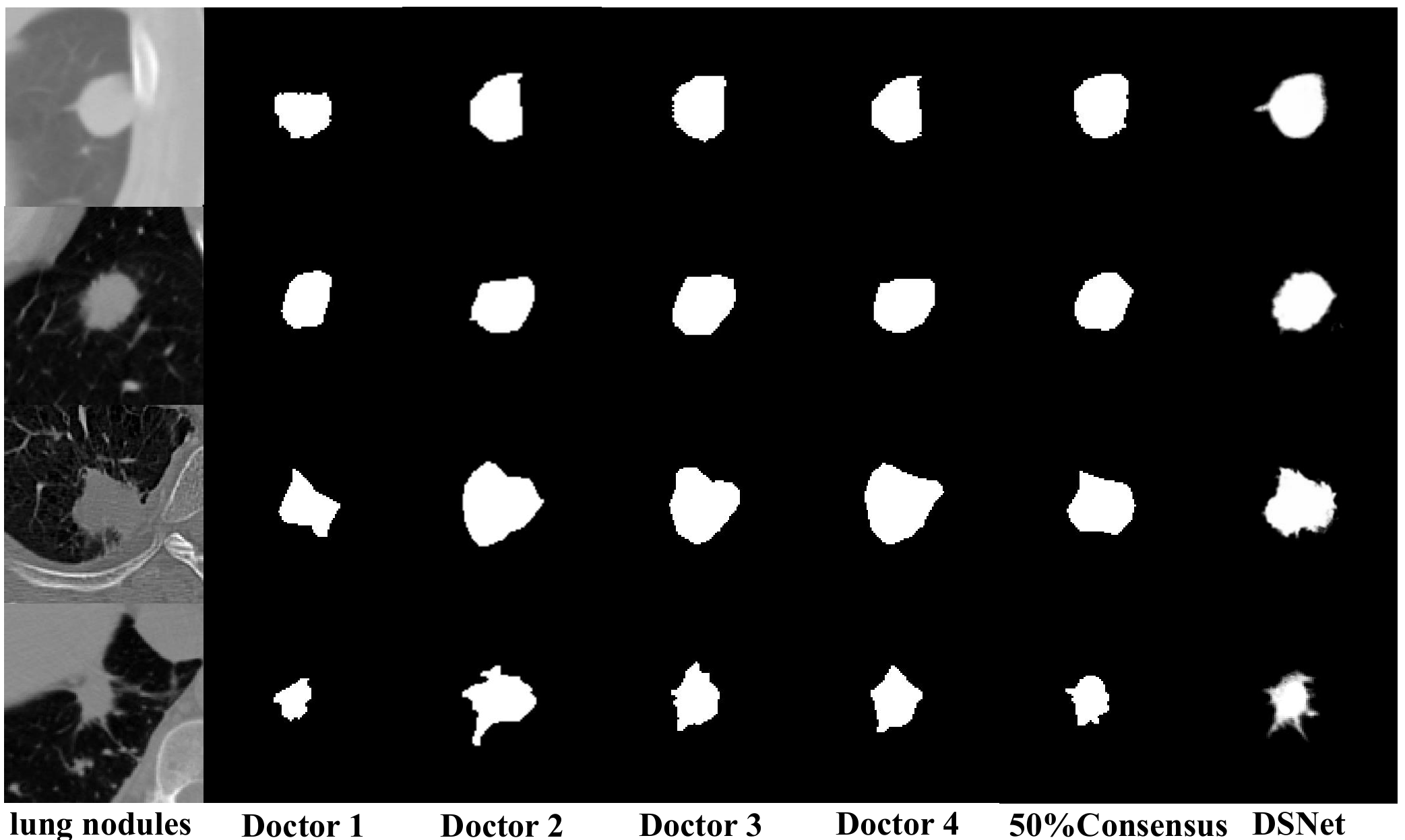}
\end{center}
   \caption{Visual comparison results between our DSNet and doctors. \textbf{Our DSNet has an amazing visual effects
that even exceeds binary ground truth (50 \% consensus).} }
\label{fig:lidc}
\end{figure}

Tab.~\ref{tab:robust} shows the comparison results of our DSNet and four human doctors on the LIDC dataset. 
The DICE between any two masks (row and column) is reported in the table.
Due to the differences and contradictions in labeling between the four radiologists, the 50 \% consensus criterion~\cite{2011Segmentation} is often used as the ground truth. It can be observed from Tab.~\ref{tab:robust} that our DSNet, even though it is trained on \textit{LNSM} dataset, is still closer to 50\% consensus than all doctors. 
On the one hand,  the results demonstrate that our DSNet has strong robustness and can deal with cross-domain challenges, which implies the potential for clinical application. On the other hand, the results mean that the \textit{LNSM} dataset we proposed is also universal and robust. 
In addition, Fig.~\ref{fig:lidc} shows the visual comparison results between our DSNet and doctors. 
Our DSNet not only has a higher DICE score but also has an impressive visual result that even exceeds ground truth (50 \% Consensus).  It’s worth mentioning that our visual results also have been approved by clinical experts.

\subsection{Generalization Study}
\label{sec: expg}
To verify the generalization of our method to other medical tasks, we perform experiments on ISBI~\cite{ISBIchallenge} and  PROMISE~\cite{PROMISE12}. The ISBI is a neuronal structure segmentation dataset and the PROMISE is a prostate MRI dataset. After data augmentation, ISBI consists of three subsets: training, verification, and testing (210, 60, and 30 samples respectively), while PROMISE consists of three subsets: training, verification, and testing ((460, 65, and 130 samples respectively). We use the \textit{Soft Mask} labeling \textbf{strategy (1)} proposed in Sec.~\ref{sec: define} to obtain the \textit{Soft Masks} for supervision. As shown in Tab.~\ref{tab:medical}, Our DSNet achieves the highest DICE and PPV scores and competitive SEN scores on these tasks that also require accurate segmentation. Our DSNet is also competent for medical segmentation tasks from cells (ISBI) to organs (PROMISE) and shows excellent generalization.

\begin{table}[thb]
\caption{Comparison of methods on ISBI and PROMISE. The \textbf{Left} column: Results of ISBI. The \textbf{Right} column: Results of PROMISE.}
\centering
\renewcommand\arraystretch{1.5} 
\setlength{\tabcolsep}{1.2mm}{
\begin{tabular}{c|c:c|c:c|c:c|c:c|c:c} 
\hline
\multicolumn{1}{c}{Methods} & \multicolumn{2}{c}{UNet} & \multicolumn{2}{c}{UNet3+} & \multicolumn{2}{c}{INFNet} & \multicolumn{2}{c}{MedT} & \multicolumn{2}{c}{DSNet}  \\ 
\hline
DICE (\%)                   & 93.5 & 87.9              & 94.3 & 91.5                & 94.5 & 90.0                & 93.7 & 91.4              & \textbf{95.9} & \textbf{92.3}                \\
SEN (\%)                    & 91.4 & 88.7              & 89.9 & 87.8                & \textbf{92.1} & 88.6                & 90.7 & \textbf{89.0}              & 90.8 & 88.6                \\
PPV (\%)                    & 91.3 & 90.4              & 94.1 & 93.3                & 94.3 & 92.3                & {94.4} & 93.1              & \textbf{94.8} & \textbf{93.5}                \\
\hline
\end{tabular}
}
\label{tab:medical}
\end{table}

% Conclusion
\subsection{Discussions}
\label{sec: discussion}
\subsubsection{Time Consumption}
One limitation of our method is that our method may introduce additional time consumption.
On the one hand, the training phase requires additional time consumption when constructing the \textit{Soft Mask} supervision. However, our annotation pipeline is automatic and efficient as reported in Sec.~\ref{sec label}. 
Our method only takes 20 $ms$ (trimap 5 $ms$, close-form-matting 15 $ms$) for a 256 $\times$ 256 image to obtain a high-quality \textit{Soft Mask} label. 
Once the \textit{Soft Mask} dataset is obtained, this part of the time consumption will no longer be burdened.
Besides, the discriminator is deprecated during the inference phase without increasing computational cost.
On the other hand, our Detailed Representation Transfer causes computational cost due to the super-resolution model.
But note that the detail representation transfer is designed to deal with low-resolution segmentation tasks, while the input size of the super-resolution model is small, so the operating efficiency of the overall pipeline will not be significantly affected.
Compared with the performance improvement, this extra calculation is worthy.

\begin{table}[H]
\centering
\caption{Inference time ablation on LNSM dataset. DRF: Detailed Representation Transfer.}
\label{tab:time}
\setlength{\tabcolsep}{0.5mm}{
\renewcommand\arraystretch{1.5} {
\begin{tabular}{lccc} 
\hline
\textit{Config.}    & \begin{tabular}[c]{@{}c@{}}LNSM \\~DICE(\%)\end{tabular} & \begin{tabular}[c]{@{}c@{}}LIDC\\~DICE(\%)\end{tabular} & Inference time ($ms$)  \\ 
\hline
w/o. DRF (SoftGAN~\cite{softgan}) & 91.63 $\pm$ 3.2                                          & 83.21 $\pm$ 7.0                                         & \textbf{5.1}           \\
w.  DRF             & \textbf{93.77 $\pm$ 3.1}                                 & \textbf{84.89 $\pm$ 7.2}                                & 6.7                    \\
\hline
\end{tabular}}}
\end{table}

Tab.~\ref{tab:time} shows our speed test results on the LNSM dataset. 
The speed is averaged on images ($64 \times 64$) with a single NVIDIA Titan V GPU.
In the future, we will explore the use of lighter super-resolution networks~\cite{yeh2021lightweight}, knowledge distillation~\cite{fu2021elastic}, and pre-trained models~\cite{shamsi2021uncertainty} to further reduce the computational cost of detail representation transfer.

\subsubsection{Relationship with Label Smooth}

Our \emph{Soft Mask} labeling can also be seen as a regularization strategy that softens the traditional one-hot type of label like label smooth in knowledge distillation~\cite{li2022distilling,zhu2020knowledge}, which can effectively suppress the over-fitting when calculating loss values. Specifically, using an independent distribution $u(y)$, the original ground truth distribution $q(y|x)$ and $u(y)$ are mixed and mapped to $q{}'(y|x)$:
\begin{align}
q{}'(y|x)=(1-\epsilon )*q(y|x)+\epsilon u(y)
\end{align}
which is consistent with our practice by using the \emph{Soft Mask} as the ground truth for training. The \emph{Soft Mask} softens the binary labels, alleviating sample imbalance and overfitting during training.
Besides, the non-binary labels of edge regions contain richer structural information like teacher model output in the knowledge distillation.

\subsubsection{Relationship with Prompt Learning in NLP}

Our \textit{Soft Mask} supervision is similar to the prompt learning~\cite{liu2021pre} that has attracted widespread attention in the natural language processing (NLP) field recently.
It can be considered that the learning of the main task (lung nodules segmentation) is improved by adding a promotion task (\textit{Soft Mask} generation).
Therefore, our method can be regarded as an innovative attempt to prompt learning in the fields of computer vision and medical image processing.

\subsubsection{Relationship with Traditional Image Processing Algorithms}
Deep learning algorithms based on neural networks have become popular in medical image processing, but traditional image processing algorithms can still bring good inspiration.
In this work, we use some traditional image processing techniques and ideas including: close-form-matting~\cite{cfm}, GrabCut~\cite{grabcut}, morphological processing~\cite{mellouli2019morphological}, and high frequency decomposition~\cite{xie2019hyperspectral}. We explore and validate the possibility and potential of incorporating these traditional algorithms in a deep learning-based segmentation framework.

\section{Conclusion}
In this work, we propose a complete solution for accurate lung nodule segmentation, including the \textit{Soft Mask} labeling pipeline and a novel DSNet, which
not only improves the network structure but also the quality of the input image and ground truth. Our method is validated to be robust and universally effective on both lung nodule segmentation and other medical datasets, and has great potential to segment other small lesions or small objects accurately while obtaining impressive visual results.

\section{Acknowledgment}
We sincerely thank the reviewers and associate editors for their helpful comments and suggestions on improving this article.

\bibliographystyle{IEEEtran}
\bibliography{ref}

%\section{Appendix}
%\includepdf[pages=-]{icme.pdf}

\end{document}